\documentclass[sigconf]{acmart}
\pdfoutput=1

\usepackage[utf8]{inputenc}
\usepackage[T1]{fontenc}
\usepackage{amsmath}
\usepackage{amsfonts}
\usepackage{amssymb}
\usepackage{amsthm}
\usepackage{dsfont}
\usepackage{array}
\usepackage{graphicx}
\usepackage{hyperref}
\usepackage{color}
\usepackage[colorinlistoftodos, prependcaption, textsize=tiny]{todonotes}

\newcommand{\E}{\operatorname{\mathbb{E}}}

\DeclareMathOperator*{\argmax}{argmax}
\newcommand{\1}{\operatorname{\mathbf{1}}}

\newcommand{\RTB}{\emph{RTB}}
\newcommand{\D}{\emph{Direct}}

\usepackage{booktabs} 

\begin{document}
\title{Optimal Allocation of Real-Time-Bidding and Direct Campaigns}

\author[G. Jauvion]{Grégoire Jauvion}
\affiliation{%
  \institution{AlephD}
  \city{Paris}
}
\email{gregoire.jauvion@gmail.com}

\author[N. Grislain]{Nicolas Grislain}
\affiliation{%
  \institution{AlephD}
  \city{Paris}
}
\email{ng@alephd.com}

\renewcommand{\shortauthors}{G. Jauvion and N. Grislain}

\begin{abstract}
In this paper, we consider the problem of optimizing the revenue a web publisher gets through real-time bidding (i.e. from ads sold in real-time auctions) and direct (i.e. from ads sold through contracts agreed in advance). We consider a setting where the publisher is able to bid in the real-time bidding auction for each impression. If it wins the auction, it chooses a direct campaign to deliver and displays the corresponding ad.

This paper presents an algorithm to build an optimal strategy for the publisher to deliver its direct campaigns while maximizing its real-time bidding revenue. The optimal strategy gives a formula to determine the publisher bid as well as a way to choose the direct campaign being delivered if the publisher bidder wins the auction, depending on the impression characteristics.

The optimal strategy can be estimated on past auctions data. The algorithm scales with the number of campaigns and the size of the dataset. This is a very important feature, as in practice a publisher may have thousands of active direct campaigns at the same time and would like to estimate an optimal strategy on billions of auctions.

The algorithm is a key component of a system which is being developed, and which will be deployed on thousands of web publishers worldwide, helping them to serve efficiently billions of ads a day to hundreds of millions of visitors.
\end{abstract}

%
%



\copyrightyear{2018} 
\acmYear{2018} 
\setcopyright{acmcopyright}
\acmConference[KDD '18]{The 24th ACM SIGKDD International Conference on Knowledge Discovery \& Data Mining}{August 19--23, 2018}{London, United Kingdom}
\acmBooktitle{KDD '18: The 24th ACM SIGKDD International Conference on Knowledge Discovery \& Data Mining, August 19--23, 2018, London, United Kingdom}
\acmPrice{15.00}
\acmDOI{10.1145/3219819.3219877}
\acmISBN{978-1-4503-5552-0/18/08}

\maketitle

\section{Introduction}

Web publishers use different channels to sell their ads. We consider in this paper two channels: \RTB{} (real-time bidding) and \D{}. In \RTB{}, the publisher sells ads through an ad exchange where buyers compete in auctions happening in real time. In \D{}, the ads are sold through direct campaigns.

A direct campaign is a contract in which a publisher commits to deliver a specified number of ads to an advertiser while ensuring that some KPIs are met (for example the publisher may guarantee a minimum number of clicks to the advertiser). Should the publisher not deliver the impressions as defined in the contract, it pays a penalty to the advertiser.

A natural way for a publisher to allocate impressions between the two channels is to bid in the \RTB{} auction for every impression: if it wins the auction, the publisher chooses a direct campaign to sell the impression, otherwise the impression is sold in the \RTB{} market.

The first section gives a review of past work related to this topic. The second section states the optimization problem to solve. The two following sections derive optimality conditions and present an algorithm to build an optimal delivering strategy. The last section presents the results of experiments realized on a real \RTB{} auctions dataset.

\section{Related work}

The problem of maximizing a web publisher advertising revenue is much studied in the literature. Because of the complexity of the online advertising ecosystem, and because the publishers sell their ads through different channels, this problem leads to diverse approaches.

\cite{ostrovsky}, \cite{yuan} and \cite{alephd} consider a publisher selling its ads through \RTB{} only, and define strategies for the publisher to set appropriate reserve prices in each auction to optimize its \RTB{} revenue. In \cite{ostrovsky}, the reserve prices are set in a static way (i.e. an optimal reserve price is estimated on a large set of auctions), whereas \cite{yuan} and \cite{alephd} define methodologies to set dynamic reserve prices which are predicted in real time before an auction happens.

\cite{budget_allocation} and \cite{budget_smoothing} study the optimal delivery of a budget across time in the \RTB{} market. This problem known as budget pacing has been much studied and presents strong similarities with the problem of delivering a direct campaign. In both articles, the bids are updated online to optimize the budget delivery.

\cite{roels} and \cite{bharadwaj} focus on \D{} revenue maximization. They consider that the impressions can be sold in direct campaigns only, and define strategies to allocate optimally a set of impressions between all campaigns in order to maximize the publisher revenue (i.e. minimize the penalty the publisher pays if some campaigns are not delivered as agreed in the contract). \cite{roels} formulates this problem as a dynamic programming problem, and \cite{bharadwaj} proposes an efficient algorithm to allocate near-optimally the impresions between the campaigns.

\cite{ghosh}, \cite{chen}, \cite{chen_2} and \cite{balseiro} study the joint optimization of the publisher revenue coming from \RTB{} and \D{}. They consider a setting similar to our setting, in which the publisher bids in auctions to compete with \RTB{} bidders.

In \cite{ghosh}, the quality of a direct campaign delivery is represented by a utility function, and the quality of each impression is linked directly to its price in the \RTB{} market, whereas we characterize a campaign delivery by some KPIs.

The model used in \cite{chen} and \cite{chen_2} is based on the modeling of the supply and the demand on an aggregated level, while we analyze data auction per auction without needing to model the supply and demand.

In \cite{balseiro}, a campaign is defined by a fixed number of impressions which is supposed to be delivered exactly, and a per-impression quality. The algorithm aims to maximize a weighted sum of the \RTB{} revenue and of the direct campaigns qualities. The main difference with our approach is that the campaigns are supposed to be completely delivered. In our approach, we enable under-delivery of a campaign if it brings a higher revenue to the publisher, which makes the problem more complex. Also, the approach described in \cite{balseiro} may be hard to scale due to the allocation plan used to allocate similar campaigns. We solve this issue by regularizing the optimization problem.

Finally, \cite{wang} and \cite{chen_thesis} present stochastic models for supply and demand, and link the problem of \D{} revenue maximization with the problem of option pricing in finance.

\section{Problem statement}

In this section, we introduce the two channels used by the publisher to sell its ads: \RTB{} and \D{}. Then, we derive the global revenue the publisher gets from both channels, and formulate the revenue optimization problem.

All random variables are defined with respect to the filtration $\mathcal{F}_{t_0}$, which is the filtration representing the information available at time $t=t_0$, i.e. before the first auction happens.

\subsection{\RTB{} market}

We represent the \RTB{} market as a set of $N$ time-ordered auctions happening at times $t_1,\ldots,t_n,\ldots,t_N$. This approach is deterministic in that we do not model the uncertainty in the auctions to come.

In auction $n$, a set of \RTB{} buyers compete to buy the ad slot and display their ad. We note respectively $B_n$ and $C_n$ the random variables corresponding to the highest bid and the second highest bid in the auction (possibly $0$ if there are not enough buyers bidding in the auction).

We note $F_n(.)$ (respectively $f_n(.)$) the marginal cumulative distribution function of $B_n$ (respectively its probability density function).

\subsection{Direct campaigns}

The \D{} channel is represented by $K$ direct campaigns $\mathcal{C}_1,\ldots,\mathcal{C}_K$ booked by the publisher at time $t=t_0$. We assume that the set of campaigns is static, i.e. existing campaigns can not be cancelled, and no new campaign is booked.

Each campaign is defined by a contractual revenue, which is the revenue the publisher is paid to deliver the campaign, a set of goals and a penalty function giving the penalty the publisher has to pay when these goals are not met.

\subsubsection{Contractual revenue of the campaign}

We note $\pi_k$ the revenue the publisher gets to deliver the campaign. It is worth noting that the publisher strategy to deliver the campaign does not depend on this amount, which is contractual and fixed before the campaign is delivered.

\subsubsection{Goals of the campaign}

Let $p_k$ be the number of goals for campaign $\mathcal{C}_k$. Each goal $i$ is defined by a targeting $\mathcal{T}_{k,i}$ and a volume $g_{k,i}$ to be delivered in the targeting $\mathcal{T}_{k,i}$.

The targeting $\mathcal{T}_{k,i}$ is defined in each auction $n$ by the Bernoulli variable $X_{n,k,i}$ equal to $1$ if impression $n$ is eligible for the targeting, and $0$ otherwise. Let $\theta_{n,k,i}$ be the parameter of the Bernoulli variable $X_{n,k,i}$, which is supposed to be known before the auction happens.

A deterministic targeting corresponds to the case where we know certainly if the ad will be in the targeting when bidding in the auction, and we have in this case $\theta_{n,k,i}=1$ or $\theta_{n,k,i}=0$. For example, targeting a website or an ad placement is a deterministic targeting.

A stochastic targeting corresponds to the case where we do not know if the ad will be in the targeting before the auction happens. In this case, $\theta_{n,k,i}$ is an estimation of the probability for the ad to be in the targeting. For example, targeting only clicked impressions is a stochastic targeting, as we do not know if the ad will be clicked or not before it has been displayed.

\subsubsection{Penalty function}

We introduce the variables $\delta_{k,1},\ldots,\delta_{k,p_k}$, where $\delta_{k,i}$ corresponds to the difference between the goal $g_{k,i}$ and the volume delivered effectively on the targeting $\mathcal{T}_{k,i}$.

For campaign $\mathcal{C}_k$, the penalty function $L_k(\delta_{k,1},\ldots,\delta_{k,p_k})$ gives the penalty the publisher has to pay when it fails to deliver the goals. At the end of the campaign, this penalty should be deduced from the contractual revenue of the campaign to get the revenue the publisher got from the campaign.

A typical penalty function would be $L_k(\delta_{k,1},\ldots,\delta_{k,p_k}) = \sum_{i=1}^{p_k} l_i \times \max (\delta_{k,i}, 0)$, corresponding to the case where the publisher pays $l_i$ for each non-delivered impression on targeting $\mathcal{T}_{k,i}$.

\subsection{Auction mechanics}

Let $a_n$ and $R_n(a_n)$ be respectively the publisher bid and the \RTB{} revenue in auction $n$. We note also $r_n(a_n)=\mathbb{E}[R_n(a_n)]$ the expected \RTB{} revenue. The auction mechanics is the following:
\begin{itemize}
\item If $a_n \geq B_n$, the publisher bidder wins the auction and the publisher chooses a direct campaign to display. We note $Q_n$ the multinoulli variable with parameters $(q_{n,1},\ldots,q_{n,K})$ such that campaign $\mathcal{C}_k$ is chosen with probability $q_{n,k}$. The \RTB{} revenue for this auction is $R_n(a_n)=0$
\item If $a_n < B_n$, the ad is sold in the \RTB{} market. When the auction mechanism is a second-price auction, which is the most usual mechanism, we have $R_n(a_n) = \mathds{1}_{a_n < B_n} \max(C_n,a_n)$\footnote{This outcome corresponds to the case where no reserve price is set in the auction. The case where a reserve price is set is handled by replacing $C_n$ by the maximum value between the second highest bid and the reserve price.}. When the auction is a first-price auction, we have $R_n(a_n) = \mathds{1}_{a_n < B_n} B_n$
\end{itemize}

\subsection{Volumes delivered on direct campaigns}

Let $V_{k,i}$ be the volume delivered on the targeting $\mathcal{T}_{k,i}$ after the $N$ auctions have occured. It can be written:
$$V_{k,i} = \sum_{n=1}^N \mathds{1}_{a_n \geq B_n} \mathds{1}_{Q_n=k} X_{n,k,i}$$

We note $v_{k,i} = \mathbb{E}[V_{k,i}]$. We assume that for each $(k,i)$, the random variables $B_n$, $Q_n$ and $X_{n,k,i}$ are independent, which gives:
$$v_{k,i} = \sum_{n=1}^N F_n(a_n)q_{n,k} \theta_{n,k,i}$$

Note that the independence hypothesis between $B_n$, $Q_n$ and $X_{n,k,i}$ is not a strong hypothesis: indeed, as these random variables are specific to auction $n$, they include already all the information we may know about the auction.

\subsection{Publisher revenue}

Let $Y$ be the total publisher revenue coming both from \RTB{} and \D{}, and let $y=\mathbb{E}[Y]$ be its expected value. It depends on the publisher bids $(a_n)_{n=1}^N$, and on the parameters $(q_{n,1},\ldots,q_{n,K})_{n=1}^N$ used to select the winning campaign when the publisher bidder wins the auction.

It writes:
\begin{equation}
\begin{split}
Y\left((a_n)_{n=1}^N,(q_{n,1},\ldots,q_{n,K})_{n=1}^N\right) =\\
\sum_{n=1}^N R_n(a_n) + \sum_{k=1}^K \left(\pi_k - L_k(V_{k,1}-g_{k,1},\ldots,V_{k,p_k}-g_{k,p_k}) \right)
\end{split}
\end{equation}

To simplify the expression of the expected revenue, we make the following approximation:
\begin{equation}
\begin{split}
\mathbb{E}[L_k(V_{k,1}-g_{k,1},\ldots,V_{k,p_k}-g_{k,p_k})] \approx\\
L_k(\mathbb{E}[V_{k,1}-g_{k,1}],\ldots,\mathbb{E}[V_{k,p_k}-g_{k,p_k}])
\end{split}
\end{equation}

This approximation tends to be well verified when the volumes delivered on all targetings are large.

The expected publisher revenue is finally written:
\begin{equation}
\begin{split}
y\left((a_n)_{n=1}^N,(q_{n,1},\ldots,q_{n,K})_{n=1}^N\right) =\\
\sum_{n=1}^N r_n(a_n) + \sum_{k=1}^K \left(\pi_k - L_k(v_{k,1}-g_{k,1},\ldots,v_{k,p_k}-g_{k,p_k}) \right)
\end{split}
\end{equation}

Note that the publisher bidder impacts the \RTB{} revenue in two ways: by removing impressions from the \RTB{} channel to deliver direct campaigns, and by modifying the \RTB{} closing price (for example when the publisher bid $a_n$ acts as a second price in a second-price auction mechanism).

\subsection{Optimization problem statement}

In this section, we note $(a_n)$ the sequence of all $a_n$ ($n=1,\ldots,N$), $(q_{n,k})$ the sequence of all $q_{n,k}$ ($n=1,\ldots,N$ and $k=1,\ldots,K$), and $(v_{k,i})$ the sequence of all $v_{k,i}$ ($k=1,\ldots,K$ and $i=1,\ldots,p_k$).

The optimization problem is written the following way:
\begin{equation}
\begin{split}
\max_{(a_n),(q_{n,k}),(v_{k,i})} \sum_{n=1}^N r_n(a_n) +\\
\sum_{k=1}^K \left( \pi_k - L_k(v_{k,1}-g_{k,1},\ldots,v_{k,p_k}-g_{k,p_k})\right)
\end{split}
\end{equation}

s.t
\begin{equation}
  \left\{
      \begin{aligned}
        \forall k,i \text{ } & v_{k,i}=\sum_{n=1}^N F_n(a_n)q_{n,k}\theta_{n,k,i} \\
        \forall n \text{ } & \sum_{k=1}^K q_{n,k}=1 \\
        \forall n,k \text{ } & q_{n,k} \geq 0 \\
      \end{aligned}
    \right.
\end{equation}

Note that the variables $(v_{k,i})$, which represent the volumes delivered on all targetings, are considered to be variables to optimize whereas they derive directly from $(a_n)$ and $(q_{n,k})$. This relation is introduced as an equality constraint. Stating the problem this way makes its analysis and the presentation of the optimization algorithm much simpler.

\section{Optimality conditions}

This section writes the Karush-Kuhn-Tucker conditions of the optimization problem stated in the previous section for two kinds of penalty functions: differentiable functions and rectified linear functions (which are not differentiable everywhere but which are widely used in practice). The lagrangian of the problem and the complete derivation of the Karush-Kuhn-Tucker conditions are given in the appendix.

\subsection{Optimality conditions when the penalty functions are differentiable}

We introduce the score of campaign $\mathcal{C}_k$ in auction $n$, defined by:
$$c_{n,k} = - \sum_{i=1}^{p_{k}} \theta_{n,k,i} \frac{\partial L_k}{\partial v_{k,i}}(v_{k,1}-g_{k,1},\ldots,v_{k,p_k}-g_{k,p_k})$$

\subsubsection{Optimality condition on $q_{n,k}$}

The optimality condition on $q_{n,k}$ writes:
$$q_{n,k}>0 \Rightarrow k \in \argmax_k c_{n,k}$$

In the particular case where there is one unique campaign $\mathcal{C}_{k_0}$ maximizing $c_{n,k}$, we have $q_{n,k}=1$ if $k=k_0$ and $q_{n,k}=0$ otherwise. If several campaigns maximize $c_{n,k}$, the optimality conditions do not give an easy way to determine the corresponding $q_{n,k}$.

\subsubsection{Optimality condition on $a_n$}

The optimality condition on $a_n$ writes:
$$r_n^{\prime}(a_n) + f_n(a_n) \max_k c_{n,k} = 0$$

It is worth noting that the expected revenue from auction $n$, should it bring $c$ if it is sold in a direct campaign, writes $r_n(a)+F_n(a)c$. Therefore, $a_n$ is the bid maximizing the expected revenue from auction $n$ if it brings $\max_k c_{n,k}$ if it is sold in a direct campaign.

The appendix gives a detailed analysis of the existence of $a_n$ verifying this condition.

\subsection{Optimality conditions when some penalty functions are rectified linear functions}

We assume in this section that the penalty function for campaign $\mathcal{C}_k$ writes $L_k(v_{k,1},\ldots,v_{k,p_k}) = \sum_{i=1}^{p_k} l_i \max(g_{k,i}-v_{k,i},0)$. This kind of penalty function is widely used in practice but is not differentiable.

The optimality conditions are similar to the ones obtained when the penalty functions are differentiable, but where the score of campaign $\mathcal{C}_k$ is written the following way:
$$
c_{n,k} = \sum_{i=1}^{p_k} \theta_{n,k,i} \kappa_{k,i} \mbox{ with } \left\{
    \begin{array}{ll}
        \kappa_{k,i} = l_i & \mbox{ if } v_{k,i}<g_{k,i} \\
        \kappa_{k,i} \in \mathopen{[} 0,l_i \mathclose{]} & \mbox{ if } v_{k,i}=g_{k,i} \\
        \kappa_{k,i} = 0 \mbox{ if } v_{k,i}>g_{k,i}
    \end{array}
\right.
$$

\section{Optimization algorithm}

This section proposes an algorithm to estimate an optimal strategy (i.e. $(a_n)$ and $(q_{n,k})$) on a \RTB{} auctions dataset. We assume that the dataset is processed in batches of auctions. In practice, the dataset is shuffled before processing it to ensure the batches are statistically equivalent.

\subsection{Description of the algorithm when the penalty functions are differentiable}

When the penalty functions are differentiable, the algorithm estimates the optimal volumes delivered on all targetings, which are the only variables needed to estimate in each auction $n$ the scores of the campaigns and then the optimal $a_n$ and $q_{n,k}$.

We note $v_{k,i}^{(j)}$ the estimation of these volumes after having processed $j$ batches. A naive initialization of the algorithm is $v_{k,i}^{(0)} = 0$, but a better initialization can accelerate significantly the convergence of the algorithm. For example, it may be relevant to initialize the algorithm with $v_{k,i}^{(0)} = g_{k,i}$.

The $j$-th batch of auctions is processed with two steps.

\subsubsection{Step 1: loop on the batch and apply the strategy determined using $v_{k,i}^{(j-1)}$}

In this section, we note $(c_{n,k}^{(j-1)})$ the scores of the campaigns in auction $n$ computed when the volumes delivered on all targetings are $(v_{k,i}^{(j-1)})$.

We note $\widehat{a_n}$ and $\widehat{q_{n,k}}$ the publisher bid and the campaigns probabilities applied in auction $n$. They are determined with the following equations, which are obtained by writing the optimality conditions when the volumes are given by $v_{k,i}^{(j-1)}$:
\begin{equation}
  \left\{
      \begin{aligned}
        \widehat{q_{n,k}}>0 \Rightarrow k \in \argmax_k c_{n,k}^{(j-1)} \\
        r_n^{\prime}(\widehat{a_n}) + f_n(\widehat{a_n}) \max_k c_{n,k}^{(j-1)} = 0 \\
      \end{aligned}
    \right.
\end{equation}

When $\argmax_k c_{n,k}^{(j-1)}$ contains one element only, it is selected, otherwise a campaign is randomly chosen among those with the maximum score. We note $\widehat{v_{k,i}}$ the volumes delivered on all campaigns when this strategy is applied on the $j$-th batch.

\subsubsection{Step 2: determine $v_{k,i}^{(j)}$}

We introduce the ratio of the number of auctions in one batch to the number of auctions in the dataset, noted $\rho$. We introduce also the learning rate of the algorithm $\alpha_j$. $v_{k,i}^{(j)}$ is determined with:
$$\forall k,i: v_{k,i}^{(j)} = v_{k,i}^{(j-1)} + \alpha_j (\widehat{v_{k,i}} - \rho \times v_{k,i}^{(j-1)})$$

This update methodology is based on the fact that at the optimum, the volume delivered on a targeting on one batch of data $\widehat{v_{k,i}}$ is equal in average to $\rho \times v_{k,i}^{(j-1)}$.

The choice of $(\alpha_j)$ has a major impact on the convergence speed of the algorithm. If $\alpha_j$ is too high, the algorithm may not converge, whereas a too low value makes the convergence very slow. We use in practice a decreasing $\alpha_j$, for example $\alpha_j \propto \frac{1}{j}$.

The size of the batches is an important parameter of the algorithm. In practice, we notice that the smaller the batches, the faster converges the algorithm, where the convergence speed is defined as the number of auctions to process to attain a given level of convergence. However, the update step performed for each batch is computationally expensive as it requires to update the volumes for all campaigns. In practice, we use batches composed of a few hundreds of auctions.

\subsection{Description of the algorithm when some penalty functions are rectified linear functions}

We assume in this section that the penalty function for campaign $\mathcal{C}_k$ writes $L_k(v_{k,1},\ldots,v_{k,p_k}) = \sum_{i=1}^{p_k} l_i \max(g_{k,i}-v_{k,i},0)$. The score $c_{n,k}$ in auction $n$ writes in this case $c_{n,k} = \sum_{i=1}^{p_k} \theta_{n,k,i} \kappa_{k,i}$.

For this campaign, the algorithm estimates the values of $\kappa_{k,i}$, which are the variables needed to estimate the score of this campaign in each auction $n$. We define $\kappa_{k,i}^{(j)}$ the estimate of $\kappa_{k,i}$ after having processed $j$ batches.

A natural way to initialize the algorithm is $\kappa_{k,i}^{(0)} = 0$, which is consistent with the initialization $v_{k,i}^{(0)}=0$. We could also initialize with $\kappa_{k,i}^{(0)} = l_i$.

Let $\widehat{\kappa_{k,i}}$ be the estimated value of $\kappa_{k,i}$ on the $j$-th batch. It is estimated by comparing the volume delivered on the batch and the volume we expect to be delivered if the goal $g_{k,i}$ is reached:
$$
\widehat{\kappa_{k,i}} = \left\{
    \begin{array}{ll}
        l_i & \mbox{ if } \widehat{v_{k,i}} < \rho \times g_{k,i} \\
        0 & \mbox{ otherwise }
    \end{array}
\right.
$$

Then, the update step is similar:
$$\forall i: \kappa_{k,i}^{(j)} = \kappa_{k,i}^{(j-1)} + \alpha_j (\widehat{\kappa_{k,i}} - \kappa_{k,i}^{(j-1)})$$

\subsection{A regularization of the optimization problem}

The application of the optimal strategy consists in picking in each auction the campaign with the highest score, where the score is computed using the volumes $(v_{k,i}^{(j)})$ after the algorithm has converged.

Such a choice of the winning campaign may lead to an unwanted behaviour of the algorithm. For example, let's consider the case of two campaigns with identical definitions. At the optimum, the volumes for these two campaigns should be identical, which means that in all auctions, both campaigns will be picked with equal probabilities. In practice, the optimal volumes estimated for these two campaigns using the algorithm may be slightly different, and in this case it is easy to see that one of the campaigns will always have a higher score than the other one, which will consequently never be picked.

This simple example illustrates that a minor error in the estimation of the optimal volumes of the campaigns may have a large impact on the estimated optimal strategy.

To overcome this issue, we have written a regularized version of the optimization problem and derived the corresponding optimality conditions. The complete derivation is given in the appendix. The regularized optimality conditions are:
$$
\left\{
    \begin{array}{l}
        q_{n,k} = \frac{ \exp( c_{n,k} \frac{F_n(a_n)}{\rho} ) } {\sum_{k'=1}^K \exp(c_{n,k'} \frac{F_n(a_n)}{\rho})} \\
        r_n^{\prime}(a_n) + f_n(a_n) \sum_{k=1}^K q_{n,k} c_{n,k} = 0
    \end{array}
\right.
$$

where $\rho$ is the regularization parameter.

It is worth noting that using this regularization in the optimization phase may also accelerate the convergence of the optimization algorithm.

\section{Experiments on an auctions dataset}

We have used in this section a real \RTB{} auctions dataset corresponding to one week of video advertising data. We have filtered auctions which led to an ad (i.e. auctions won by a buyer and for which a video ad has been delivered effectively). The dataset has been shuffled before running the optimization algorithm to ensure that the batches are statistically equivalent.

In each auction, we will use the following informations:
\begin{itemize}
\item The impression id, which is a unique identifier of the impression
\item The placement id, which is a unique identifier of the ad slot. The dataset contains 4 different ad placements, called $\mathcal{P}_1$, $\mathcal{P}_2$, $\mathcal{P}_3$ and $\mathcal{P}_4$
\item The highest bid of the auction (which is strictly positive because we filtered on auctions won by a buyer)
\item We know if the impression has been viewed\footnote{The IAB viewability measure is used: an ad is considered to be viewed if at least 50\% of its surface has been within the visible area of the browser window during at least 1 second.} or clicked
\end{itemize}

Two experiments have been performed:
\begin{itemize}
\item The first experiment runs the optimization algorithm with a few campaigns defined manually, and gives an interpretation of the optimal strategy
\item The second experiment runs the optimization algorithm with many campaigns and shows that the algorithm scales with the number of campaigns
\end{itemize}

We assume that the auction mechanism is a first-price auction mechanism. This makes the optimality condition on $a$ easier to solve because it does not need any hypothesis on the highest \RTB{} bid distribution.

The optimization algorithm is run with batches of size $1000$ and with $\alpha_t = \frac{1}{t}$. The optimality conditions have been regularized with the temperature parameter $\rho = 0.5$, both when determining the optimal strategy and when applying the optimal strategy on the dataset. The algorithm is initialized with $\kappa_{k,i}=0$.

We define in this section the adjusted revenue as the \RTB{} revenue reduced by the penalty the publisher has to pay on the direct campaigns in case of under-delivery. In both experiments, we plot the adjusted revenue of the publisher when applying the estimated optimal strategy on the whole dataset, as a function of the number of batches of size $1000$ processed by the algorithm.

\subsection{Descriptive analysis of the dataset}

The table below gives the main characteristics of the dataset, per placement and globally.

\begin{center}
\begin{tabular}{c c c c c}
\hline
 & Nb Imps & Avg \RTB{} bid & Avg view rate & Avg click rate \\ \hline
$\mathcal{P}_1$ & 15.46 mlns & 13.15 \$CPM & 77.0 \% & 0.13 \% \\
$\mathcal{P}_2$ & 5.56 mlns & 15.36 \$CPM & 77.9 \% & 0.23 \% \\
$\mathcal{P}_3$ & 2.43 mlns & 15.55 \$CPM & 59.7 \% & 2.03 \% \\
$\mathcal{P}_4$ & 0.14 mlns & 25.37 \$CPM & 42.7 \% & 0.00 \% \\ \hline
Total & 23.59 mlns & 13.99 \$CPM & 75.3 \% & 0.35 \% \\ \hline
\end{tabular}
\end{center}

\subsection{First experiment: optimization with a few campaigns}

In this section, we define a direct campaign by the following sextuplet:
$$(N_{imp}, N_{view}, N_{click}, l_{imp}, l_{view}, l_{click})$$
, where $N_*$ and $l_*$ correspond to the goal and to a penalty parameter associated to the number of impressions, the number of impressions viewed and the number of impressions clicked. We assume that all campaigns target all the placements: this hypothesis makes the results simpler to interpret.

We assume that the penalty functions are rectified linear functions where $l_*$ is the penalty per non-delivered impression. If $v_{imp}$, $v_{view}$ and $v_{click}$ are the volumes of impressions, of impressions viewed and of impressions clicked which have been delivered, the penalty function writes:
$$L(v_{imp},v_{view},v_{click}) = \sum_{i \in \{imp,view,click\}} l_i \times \max(N_i-v_i,0)$$

This experiment runs the optimization algorithm with the following campaigns:
\begin{itemize}
\item 3 impressions campaigns:
  \begin{itemize}
    \item $\mathcal{C}_1=(3.10^6,.,.,5,.,.)$
    \item $\mathcal{C}_2=(3.10^6,.,.,10,.,.)$
    \item $\mathcal{C}_3=(3.10^6,.,.,20,.,.)$
  \end{itemize}
\item 3 viewability campaigns:
  \begin{itemize}
    \item $\mathcal{C}_4=(.,2.10^6,.,.,5,.)$
    \item $\mathcal{C}_5=(.,2.10^6,.,.,15,.)$
    \item $\mathcal{C}_6=(.,2.10^6,.,.,30,.)$
  \end{itemize}
\item 3 click campaigns:
  \begin{itemize}
    \item $\mathcal{C}_7=(.,.,2.10^3,.,.,200)$
    \item $\mathcal{C}_8=(.,.,2.10^3,.,.,500)$
    \item $\mathcal{C}_9=(.,.,2.10^3,.,.,1000)$
  \end{itemize}
\end{itemize}

To run the experiment, we have used simple viewability and click predictors which predict the average viewability rate and click rate of the placement. These predictors are used before each auction to compute the scores of the campaigns. Using more advanced predictors based on more features would clearly improve the results.

\begin{figure}
\centering
\includegraphics[scale=0.35]{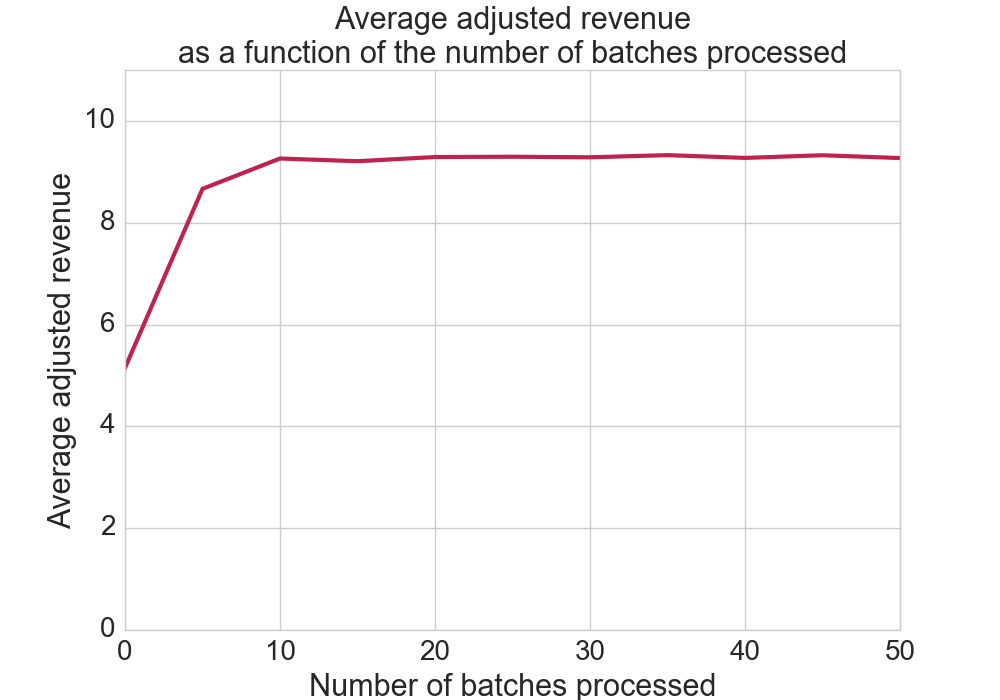}
\caption{First experiment: average adjusted revenue as a function of the number of batches processed}
\label{experiment_1}
\end{figure}

Figure \ref{experiment_1} plots the adjusted revenue estimated on the whole dataset as a function of the number of batches processed by the optimization algorithm. The intialization point corresponds to the case where no impression is sold in a direct campaign and it is very far from the optimum in terms of adjusted revenue. The algorithm converges after having processed a few dozens of batches.

The table below has been built by applying the optimal strategy estimated after having processed $100$ batches of auctions. It gives, for each campaign, the number of impressions delivered on each placement, as a percentage of the goal of the campaign. It gives also the undelivered volume (i.e. the goal less the volume delivered) as a percentage of the goal of the campaign.

\begin{center}
\begin{tabular}{c | c c c c | c}
\hline
Campaign & P1 & P2 & P3 & P4 & Undelivered \\ \hline
$\mathcal{C}_1$ & 12\% & 4\% & 1\% & 0\% & 83\% \\
$\mathcal{C}_2$ & 49\% & 15\% & 4\% & 0\% & 32\% \\
$\mathcal{C}_3$ & 73\% & 23\% & 5\% & 0\% & 0\% \\ \hline
$\mathcal{C}_4$ & 8\% & 3\% & 0\% & 0\% & 89\% \\
$\mathcal{C}_5$ & 70\% & 22\% & 2\% & 0\% & 6\% \\
$\mathcal{C}_6$ & 75\% & 23\% & 2\% & 0\% & 0\% \\ \hline
$\mathcal{C}_7$ & 0\% & 0\% & 19\% & 0\% & 81\% \\
$\mathcal{C}_8$ & 0\% & 0\% & 83\% & 0\% & 17\% \\
$\mathcal{C}_9$ & 0\% & 0\% & 100\% & 0\% & 0\% \\ \hline
\end{tabular}
\end{center}

We see that the campaigns $\mathcal{C}_1$, $\mathcal{C}_2$, $\mathcal{C}_4$, $\mathcal{C}_5$, $\mathcal{C}_7$ and $\mathcal{C}_8$ are not fully delivered: for these campaigns, the penalty in case of under-delivery is low compared to the opportunity cost incurred by the publisher by not selling impressions on the \RTB{} market. Also, the campaigns targeting clicks are mainly delivered on placement $\mathcal{P}_3$, which has a significantly higher click probability whereas the \RTB{} bid is not significantly higher than on the other placements.

\subsection{Second experiment: optimization with many campaigns}

We note in this section $U(a,b)$ the uniform distribution on $[a,b]$. In this experiment, we run the optimization with $K$ campaigns, and each campaign is generated the following way:
\begin{itemize}
\item It targets all impressions where $h(\text{imp\_id}) \equiv 1 \mod M$, where $h$ is a hash function, $\text{imp\_id}$ is the id of the impression, and $M$ is drawn from $U(10,100)$. This methodology ensures that the targetings of a pair of campaigns has most of the time a non-empty intersection (which makes probably the optimization harder, as each impression must be allocated efficiently among all the campaigns targeting it)
\item The impression goal $N_{imp}$ is drawn from the distribution $U(0,N)$, where $N$ is chosen such that the sum of the goals on all campaigns represent in average $40\%$ of the total number of impressions available. Consequently, $N$ decreases when $K$ increases. Note that in this experiment, we do not define any goals on the number of impressions clicked or viewed
\item The penalty parameter $l_{imp}$ associated to the impressions goal is drawn from the distribution $U(0,50)$
\end{itemize}

\begin{figure}
\centering
\includegraphics[scale=0.35]{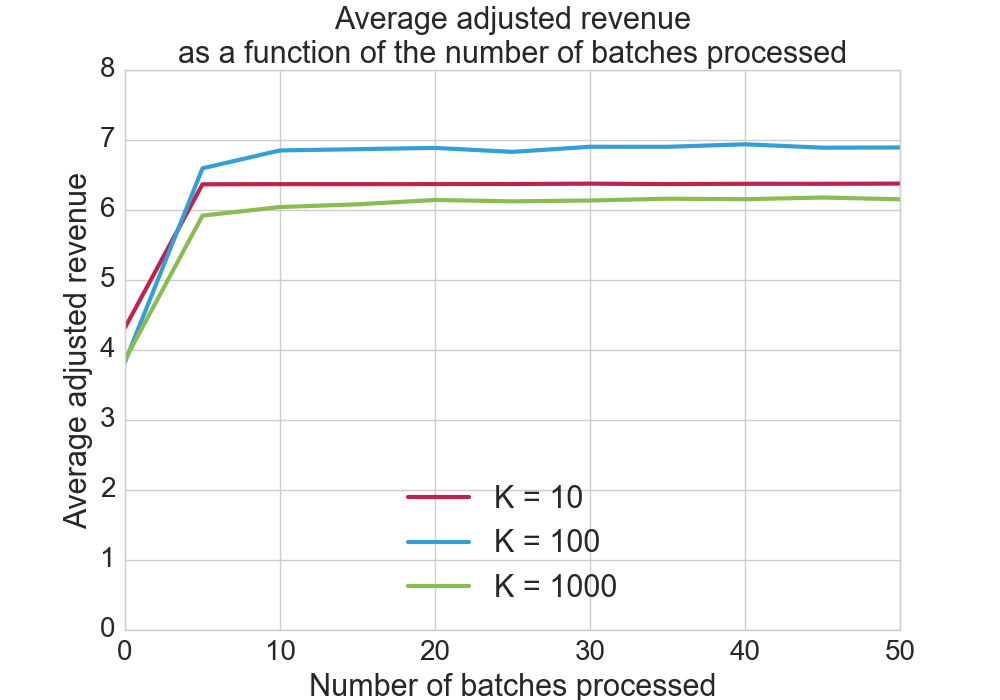}
\caption{Second experiment: average adjusted revenue as a function of the number of batches processed}
\label{experiment_2}
\end{figure}

Figure \ref{experiment_2} plots the adjusted revenue of the publisher on the whole dataset as a function of the number of batches processed by the optimization algorithm, for different values of the number of campaigns $K$.

We see on the figure that increasing the number of campaigns does not slow the convergence of the algorithm, and a few dozens of batches are enough to converge to the optimum. In practice, the optimum is attained in less than a few minutes with $1000$ campaigns.

\section{Conclusions and future directions}

In this paper, we study the problem of optimizing the revenue a web publisher gets through \RTB{} and \D{}. For each impression, the publisher bids in the \RTB{} auction and chooses the direct campaign to deliver should it win the auction. We have been able to define an algorithm to determine an optimal strategy for the publisher in this setting.

The experiments performed on a \RTB{} auctions dataset show that the algorithm scales with the number of campaigns, as increasing the number of campaigns does not slow the convergence. With $1000$ campaigns, an optimal strategy can be estimated in a few minutes.

Also, applying the optimal strategy is particularly simple and the most computationally intensive part is to compute the score of each campaign, which is a weighted sum of the predictors $\theta_{n,k,i}$. In practice, the publisher has a few milliseconds to determine its bid. Should the number of campaigns be too high to be able to compute all the scores, the publisher should define a strategy to sample the campaigns considered for a given impression.

In practice, a web publisher should estimate the optimal strategy on a long period of time (ideally on several months) to make it as stable as possible. Applying it in the future relies on the hypothesis that the future \RTB{} market is similar to the past auctions used in the estimation, in terms of volume and bid level. The estimation can also be adjusted to reflect an expected change in the \RTB{} market.

When the \RTB{} market characteristics differ too much from the dataset used in the estimation, the publisher should define a methodology to adjust the strategy. Such a methodology can be based for example on a controller which updates the strategy from time to time. This is a different topic which has been studied specifically in the literature (see \cite{brendan}, \cite{niklas}).

The algorithm presented in the paper combined with a controller form a system which is being developed to be deployed on thousands of publishers worldwide.

Finally, the hypothesis that the set of campaigns is known when estimating the strategy is well suited to the theoretical analysis of the problem and to the derivation of an estimation algorithm. In practice, the publisher should make hypotheses on the campaigns which will be agreed in the future.

\section{Appendix}

\subsection{Derivation of the optimality conditions}

The lagrangian of the problem writes:
\begin{equation}
\begin{split}
\mathcal{L}((a_n),(q_{n,k}),(v_{k,i})) = \sum_{n=1}^N r_n(a_n) +\\
\sum_{k=1}^K \left( \pi_k - L_k(v_{k,1}-g_{k,1},\ldots,v_{k,p_k}-g_{k,p_k}) \right) \\
- \sum_{k=1}^K \sum_{i=1}^{p_k} \lambda_{k,i}(v_{k,i}-\sum_{n=1}^N F_n(a_n)q_{n,k}\theta_{n,k,i}) \\
- \sum_{n=1}^N \mu_{n} (\sum_{k=1}^K q_{n,k}-1) + \sum_{n=1}^N \sum_{k=1}^K \eta_{n,k} q_{n,k}
\end{split}
\end{equation}

where $\lambda_{k,i}$, $\mu_{n}$ and $\eta_{n,k}$ are the lagrange multipliers corresponding to the constraints.

The Karush-Kuhn-Tucker optimality conditions are:
\begin{equation}
  \left\{
      \begin{aligned}
        \forall n, \frac{\partial \mathcal{L}}{\partial a_n} & = & r_n^{\prime}(a_n) + f_n(a_n) \sum_{k=1}^K q_{n,k} (\sum_{i=1}^{p_k} \lambda_{k,i} \theta_{n,k,i})=0\\
        \forall n, k, \frac{\partial \mathcal{L}}{\partial q_{n,k}} & = & F_n(a_n) \sum_{i=1}^{p_k} \lambda_{k,i} \theta_{n,k,i} - \mu_n + \eta_{n,k}=0  \\
        \forall k, i, \frac{\partial \mathcal{L}}{\partial v_{k,i}} & = & - \frac{\partial L_k}{\partial v_{k,i}}(v_{k,1}-g_{k,1},\ldots,v_{k,p_k}-g_{k,p_k}) - \lambda_{k,i} = 0 \\
      \end{aligned}
    \right.
\end{equation}

\subsubsection{Conditions on $q_{n,k}$}

From the second equation we get for auction $n$:
$$\forall k, F_n(a_n) \sum_{i=1}^{p_k} \lambda_{k,i} \theta_{n,k,i} = \mu_n-\eta_{n,k}$$

Let's consider the set $M_n$ of $k$-s maximizing $\sum_{i=1}^{p_k} \lambda_{k,i} \theta_{n,k,i}$:
$$M_n = \argmax_k \sum_{i=1}^{p_k} \lambda_{k,i} \theta_{n,k,i}$$

If $F_n(a_n)>0$, we get:
$$M_n = \argmax_k (\mu_n-\eta_{n,k})$$

As $\eta_{n,k}$ is the Lagrange multiplier corresponding to the constraint $q_{n,k} \geq 0$, we have $\forall k, \min(\eta_{n,k},q_{n,k})=0$. We know also that there must be at least one $k$ such that $\eta_{n,k}=0$ (otherwise all $q_{n,k}$ would be equal to $0$ and they couldn't sum to $1$).

We deduce that $\max_k (\mu_n-\eta_{n,k}) = \mu_n$, and finally: if $k \in M_n$, $\eta_{n,k}=0$ (and $q_{n,k}>0$), otherwise $\eta_{n,k}>0$ (and $q_{n,k}=0$).

We introduce the score of campaign $\mathcal{C}_k$ in auction $n$, defined by:
$$c_{n,k} = - \sum_{i=1}^{p_{k}} \theta_{n,k,i} \frac{\partial L_k}{\partial v_{k,i}}(v_{k,1}-g_{k,1},\ldots,v_{k,p_k}-g_{k,p_k})$$

We obtain the following optimality condition:
$$q_{n,k}>0 \Rightarrow k \in \argmax_k c_{n,k}$$

\subsubsection{Conditions on $a_n$}

The first equation gives the optimality condition on $a_n$:
$$r_n^{\prime}(a_n) + f_n(a_n) \max_k c_{n,k} = 0$$

\subsection{Particular case where some penalty functions are rectified linear functions}

We assume in this section that the penalty function for campaign $\mathcal{C}_k$ writes $L_k(v_{k,1},\ldots,v_{k,p_k}) = \sum_{i=1}^{p_k} l_i \max(g_{k,i}-v_{k,i},0)$. This kind of penalty function is widely used in practice but is not differentiable.

Let $\partial \mathcal{L}_{v_{k,i}}$ be the sub-gradient of the lagrangian $\mathcal{L}$ with respect to $v_{k,i}$, which writes:
$$
\partial \mathcal{L}_{v_{k,i}} = \left\{
    \begin{array}{ll}
        \{ -l_i-\lambda_{k,i} \} & \mbox{ if } v_{k,i}<g_{k,i} \\
        \mathopen{[} -l_i-\lambda_{k,i},-\lambda_{k,i} \mathclose{]} & \mbox{ if } v_{k,i}=g_{k,i} \\
        \{ -\lambda_{k,i} \} & \mbox{ if } v_{k,i}>g_{k,i}
    \end{array}
\right.
$$

The Karush-Kuhn-Tucker optimality condition with respect to $v_{k,i}$ writes: $0 \in \partial \mathcal{L}_{v_{k,i}}$. This gives similar optimality conditions where the score of campaign $\mathcal{C}_k$ is written the following way:
$$
c_{n,k} = \sum_{i=1}^{p_k} \theta_{n,k,i} \kappa_{k,i} \mbox{ with } \left\{
    \begin{array}{ll}
        \kappa_{k,i} = l_i & \mbox{ if } v_{k,i}<g_{k,i} \\
        \kappa_{k,i} \in \mathopen{[} 0,l_i \mathclose{]} & \mbox{ if } v_{k,i}=g_{k,i} \\
        \kappa_{k,i} = 0 \mbox{ if } v_{k,i}>g_{k,i}
    \end{array}
\right.
$$

\subsection{Analysis of the optimality condition on $a$}

In this section, we note $F_b$ and $F_c$ the cumulative probability functions of the highest bid $B$ and the second highest bid $C$, and $f_b$ and $f_c$ their probability density functions. We assume that these functions are continuous on $[0, +\infty[$ and that $\forall a, f_b(a)>0$.

The optimality condition on $a$ states that the optimal value verifies $\frac{r^{\prime}}{f_b}(a) = -c$, where $c \geq 0$ is the maximum score of the direct campaigns.

We show in this section that in a second-price auction mechanism, this optimality condition has at least one solution under a usual hypothesis on the distributions $B$ and $C$. The solution is very simple in a first-price auction mechanism.

\subsubsection{Second-price auction mechanism}

We have:
$$r(a) = \E[\1_{a \leq B} max(a,C)] = \E[\1_{a \leq C} C] + \E[\1_{C \leq a \leq B} a]$$

Note that $\E[\1_{C \leq a \leq B}] = P(a \leq B) - P(a \leq C) = F_c(a) - F_b(a)$, because $C \leq B$.

Then:
\begin{eqnarray*}
r'(a) & = & F_c(a) - F_b(a) + a (f_c(a) - f_b(a)) - a f_c(a) \\
      & = & - a f_b(a) + (F_c(a) - F_b(a)) \\
\end{eqnarray*}

And:
$$\frac{r'}{f_b}(a) = -a + \frac{F_c(a) - F_b(a)}{f_b(a)}$$

It is worth noting that this function depends on the marginal distributions $B$ and $C$ only. It does not depend on the joint distribution $(B,C)$, which makes the analysis much simpler.

It is easy to see that $\frac{r'}{f_b}(0) = 0$ (as $F_b(0)=F_c(0)=0$). To analyze the behaviour of this function when $a \to +\infty$, we write the function $\frac{r'}{f_b}$ using the hazard rates of $B$ and $C$, $h_b = \frac{f_b}{1-F_b}$ and $h_c = \frac{f_c}{1-F_c}$:
$$\frac{r'}{f_b}(a) = -a + \frac{1}{h_b(a)} - \frac{1}{h_c(a)} \frac{f_c(a)}{f_b(a)} \leq -a + \frac{1}{h_b(a)}$$

We assume also that $\frac{1}{h_b(a)} = o(a)$. This hypothesis is quite usual and is verified by the normal and the log-normal distributions. Therefore, we have:
$$\lim_{a \to +\infty} \frac{r'}{f_b}(a) = -\infty$$

As $\frac{r'}{f_b}$ is a continuous function on $[0,+\infty[$, we conclude that under the hypothesis $\frac{1}{h_b(a)} = o(a)$, $\frac{r'}{f_b}$ is a surjective function to $[0,-\infty[$. This means that the optimality condition $\frac{r^{\prime}}{f_b}(a) = -c$ has at least one solution for every $c \geq 0$. However, we have no guarantee on the unicity of the solution.

\subsubsection{First-price auction mechanism}

The first-price auction can be modeled with $C=B$, and in this case we have $\frac{r'}{f_b}(a) = -a$. The optimality condition has one unique solution $a=c$.

\subsection{A regularization of the optimization problem}

We introduce in this section a regularization of the optimization problem which makes the optimality conditions more stable. For each auction $n$, we add the term $\rho H(Q_n)$ to the objective function, where $\rho$ is the regularization parameter which does not depend on $n$ and $H(Q_n)$ is the entropy of the distribution $Q_n$. The case $\rho \to 0$ corresponds to the non regularized optimization problem, and in the case $\rho \to +\infty$, the optimal $Q_n$ distribution is the uniform distribution.

The objective function of the problem becomes:
\begin{equation}
\begin{split}
y_{regularized}\left((a_n)_{n=1}^N,(q_{n,1},\ldots,q_{n,K})_{n=1}^N,\rho\right) =\\
\sum_{n=1}^N r_n(a_n) + \sum_{k=1}^K \left(\pi_k - L_k(v_{k,1}-g_{k,1},\ldots,v_{k,p_k}-g_{k,p_k}) \right) \\
- \rho \sum_{n=1}^N \left( \sum_{k=1}^K q_{n,k} \ln(q_{n,k}) \right)
\end{split}
\end{equation}

The Karush-Kuhn-Tucker optimality conditions write:
\begin{equation}
  \left\{
      \begin{aligned}
        \forall n, \frac{\partial \mathcal{L}}{\partial a_n} & = & r_n^{\prime}(a_n) + f_n(a_n) \sum_{k=1}^K q_{n,k} c_{n,k} = 0\\
        \forall n, k, \frac{\partial \mathcal{L}}{\partial q_{n,k}} & = & F_n(a_n) c_{n,k} - \mu_n - \rho (1 + \ln(q_{n,k})) =0  \\
        \forall i, k, \frac{\partial \mathcal{L}}{\partial v_{k,i}} & = & - \frac{\partial L_k}{\partial v_{k,i}}(v_{k,1}-g_{k,1},\ldots,v_{k,p_k}-g_{k,p_k}) - \lambda_{k,i} = 0 \\
      \end{aligned}
    \right.
\end{equation}

The second equation gives:
$$q_{n,k} = e^{-(1 + \frac{\mu_n}{\rho})} e^{c_{n,k} \frac{F_n(a_n)}{\rho}}$$

And the constraint $\sum_{k=1}^K q_{n,k} = 1$ gives:
$$q_{n,k} = \frac{ \exp( c_{n,k} \frac{F_n(a_n)}{\rho} ) } {\sum_{k'=1}^K \exp(c_{n,k'} \frac{F_n(a_n)}{\rho})}$$

Finally, the optimality conditions of the problem are the following:
\begin{itemize}
\item The $q_{n,k}$ define a Boltzmann distribution with the weights $e^{c_{n,k}}$ and a temperature $\frac{\rho}{F_n(a_n)}$
\item $a_n$ verifies $r_n^{\prime}(a_n) + f_n(a_n) \sum_{k=1}^K q_{n,k} c_{n,k} = 0$
\end{itemize}

\bigskip


\bibliographystyle{ACM-Reference-Format}
\bibliography{deterministic_holistic_yield_kdd}


\begin{thebibliography}{15}


\ifx \showCODEN    \undefined \def \showCODEN     #1{\unskip}     \fi
\ifx \showDOI      \undefined \def \showDOI       #1{#1}\fi
\ifx \showISBNx    \undefined \def \showISBNx     #1{\unskip}     \fi
\ifx \showISBNxiii \undefined \def \showISBNxiii  #1{\unskip}     \fi
\ifx \showISSN     \undefined \def \showISSN      #1{\unskip}     \fi
\ifx \showLCCN     \undefined \def \showLCCN      #1{\unskip}     \fi
\ifx \shownote     \undefined \def \shownote      #1{#1}          \fi
\ifx \showarticletitle \undefined \def \showarticletitle #1{#1}   \fi
\ifx \showURL      \undefined \def \showURL       {\relax}        \fi
\providecommand\bibfield[2]{#2}
\providecommand\bibinfo[2]{#2}
\providecommand\natexlab[1]{#1}
\providecommand\showeprint[2][]{arXiv:#2}

\bibitem[\protect\citeauthoryear{Balseiro, Feldman, Mirrokni, and
  Muthukrishnan}{Balseiro et~al\mbox{.}}{2011}]%
        {balseiro}
\bibfield{author}{\bibinfo{person}{Santiago Balseiro}, \bibinfo{person}{Jon
  Feldman}, \bibinfo{person}{Vahab Mirrokni}, {and} \bibinfo{person}{S.
  Muthukrishnan}.} \bibinfo{year}{2011}\natexlab{}.
\newblock \showarticletitle{Yield Optimization of Display Advertising with Ad
  Exchange}. In \bibinfo{booktitle}{\emph{Proceedings of the 12th ACM
  Conference on Electronic Commerce}} \emph{(\bibinfo{series}{EC '11})}.
  \bibinfo{publisher}{ACM}, \bibinfo{address}{New York, NY, USA},
  \bibinfo{pages}{27--28}.
\newblock
\showISBNx{978-1-4503-0261-6}
\urldef\tempurl%
\url{https://doi.org/10.1145/1993574.1993580}
\showDOI{\tempurl}


\bibitem[\protect\citeauthoryear{Bharadwaj, Chen, Ma, Nagarajan, Tomlin,
  Vassilvitskii, Vee, and Yang}{Bharadwaj et~al\mbox{.}}{2012}]%
        {bharadwaj}
\bibfield{author}{\bibinfo{person}{Vijay Bharadwaj}, \bibinfo{person}{Peiji
  Chen}, \bibinfo{person}{Wenjing Ma}, \bibinfo{person}{Chandrashekhar
  Nagarajan}, \bibinfo{person}{John Tomlin}, \bibinfo{person}{Sergei
  Vassilvitskii}, \bibinfo{person}{Erik Vee}, {and} \bibinfo{person}{Jian
  Yang}.} \bibinfo{year}{2012}\natexlab{}.
\newblock \showarticletitle{SHALE: An Efficient Algorithm for Allocation of
  Guaranteed Display Advertising}. In \bibinfo{booktitle}{\emph{Proceedings of
  the 18th ACM SIGKDD International Conference on Knowledge Discovery and Data
  Mining}} \emph{(\bibinfo{series}{KDD '12})}. \bibinfo{publisher}{ACM},
  \bibinfo{address}{New York, NY, USA}, \bibinfo{pages}{1195--1203}.
\newblock
\showISBNx{978-1-4503-1462-6}
\urldef\tempurl%
\url{https://doi.org/10.1145/2339530.2339718}
\showDOI{\tempurl}


\bibitem[\protect\citeauthoryear{Chahuara, Grislain, Jauvion, and
  Renders}{Chahuara et~al\mbox{.}}{2017}]%
        {alephd}
\bibfield{author}{\bibinfo{person}{Pedro Chahuara}, \bibinfo{person}{Nicolas
  Grislain}, \bibinfo{person}{Gregoire Jauvion}, {and}
  \bibinfo{person}{Jean-Michel Renders}.} \bibinfo{year}{2017}\natexlab{}.
\newblock \showarticletitle{Real-Time Optimization of Web Publisher RTB
  Revenues}. In \bibinfo{booktitle}{\emph{Proceedings of the 23rd ACM SIGKDD
  International Conference on Knowledge Discovery and Data Mining}}
  \emph{(\bibinfo{series}{KDD '17})}. \bibinfo{publisher}{ACM},
  \bibinfo{address}{New York, NY, USA}, \bibinfo{pages}{1743--1751}.
\newblock
\showISBNx{978-1-4503-4887-4}
\urldef\tempurl%
\url{https://doi.org/10.1145/3097983.3098150}
\showDOI{\tempurl}


\bibitem[\protect\citeauthoryear{{Chen}}{{Chen}}{2016a}]%
        {chen_thesis}
\bibfield{author}{\bibinfo{person}{B. {Chen}}.}
  \bibinfo{year}{2016}\natexlab{a}.
\newblock \showarticletitle{{PhD thesis, Financial Methods for Online
  Advertising}}.
\newblock  (\bibinfo{year}{2016}).
\newblock


\bibitem[\protect\citeauthoryear{{Chen}}{{Chen}}{2016b}]%
        {chen_2}
\bibfield{author}{\bibinfo{person}{B. {Chen}}.}
  \bibinfo{year}{2016}\natexlab{b}.
\newblock \showarticletitle{{Risk-Aware Dynamic Reserve Prices of Programmatic
  Guarantee in Display Advertising}}.
\newblock  (\bibinfo{year}{2016}).
\newblock


\bibitem[\protect\citeauthoryear{Chen, Yuan, and Wang}{Chen
  et~al\mbox{.}}{2014}]%
        {chen}
\bibfield{author}{\bibinfo{person}{Bowei Chen}, \bibinfo{person}{Shuai Yuan},
  {and} \bibinfo{person}{Jun Wang}.} \bibinfo{year}{2014}\natexlab{}.
\newblock \showarticletitle{A Dynamic Pricing Model for Unifying Programmatic
  Guarantee and Real-Time Bidding in Display Advertising}. In
  \bibinfo{booktitle}{\emph{Proceedings of the Eighth International Workshop on
  Data Mining for Online Advertising}} \emph{(\bibinfo{series}{ADKDD'14})}.
  \bibinfo{publisher}{ACM}, \bibinfo{address}{New York, NY, USA}, Article
  \bibinfo{articleno}{1}, \bibinfo{numpages}{9}~pages.
\newblock
\showISBNx{978-1-4503-2999-6}
\urldef\tempurl%
\url{https://doi.org/10.1145/2648584.2648585}
\showDOI{\tempurl}


\bibitem[\protect\citeauthoryear{{Ghosh}, {McAfee}, {Papineni}, and S.}{{Ghosh}
  et~al\mbox{.}}{2009}]%
        {ghosh}
\bibfield{author}{\bibinfo{person}{A. {Ghosh}}, \bibinfo{person}{P. {McAfee}},
  \bibinfo{person}{K. {Papineni}}, {and} \bibinfo{person}{{Vassilvitskii} S.}}
  \bibinfo{year}{2009}\natexlab{}.
\newblock \showarticletitle{{Bidding for Representative Allocations for Display
  Advertising}}.
\newblock  (\bibinfo{year}{2009}).
\newblock


\bibitem[\protect\citeauthoryear{{Karlsson} and {Zhang}}{{Karlsson} and
  {Zhang}}{2013}]%
        {niklas}
\bibfield{author}{\bibinfo{person}{N. {Karlsson}} {and} \bibinfo{person}{J.
  {Zhang}}.} \bibinfo{year}{2013}\natexlab{}.
\newblock \showarticletitle{{Applications of feedback control in online
  advertising}}.
\newblock  (\bibinfo{year}{2013}).
\newblock


\bibitem[\protect\citeauthoryear{Kitts, Krishnan, Yadav, Zeng, Badeau, Potter,
  Tolkachov, Thornburg, and Janga}{Kitts et~al\mbox{.}}{2017}]%
        {brendan}
\bibfield{author}{\bibinfo{person}{Brendan Kitts}, \bibinfo{person}{Michael
  Krishnan}, \bibinfo{person}{Ishadutta Yadav}, \bibinfo{person}{Yongbo Zeng},
  \bibinfo{person}{Garrett Badeau}, \bibinfo{person}{Andrew Potter},
  \bibinfo{person}{Sergey Tolkachov}, \bibinfo{person}{Ethan Thornburg}, {and}
  \bibinfo{person}{Satyanarayana~Reddy Janga}.}
  \bibinfo{year}{2017}\natexlab{}.
\newblock \showarticletitle{Ad Serving with Multiple KPIs}. In
  \bibinfo{booktitle}{\emph{Proceedings of the 23rd ACM SIGKDD International
  Conference on Knowledge Discovery and Data Mining}}
  \emph{(\bibinfo{series}{KDD '17})}. \bibinfo{publisher}{ACM},
  \bibinfo{address}{New York, NY, USA}, \bibinfo{pages}{1853--1861}.
\newblock
\showISBNx{978-1-4503-4887-4}
\urldef\tempurl%
\url{https://doi.org/10.1145/3097983.3098085}
\showDOI{\tempurl}


\bibitem[\protect\citeauthoryear{Lee, Jalali, and Dasdan}{Lee
  et~al\mbox{.}}{2013}]%
        {budget_smoothing}
\bibfield{author}{\bibinfo{person}{Kuang-Chih Lee}, \bibinfo{person}{Ali
  Jalali}, {and} \bibinfo{person}{Ali Dasdan}.}
  \bibinfo{year}{2013}\natexlab{}.
\newblock \showarticletitle{Real Time Bid Optimization with Smooth Budget
  Delivery in Online Advertising}. In \bibinfo{booktitle}{\emph{Proceedings of
  the Seventh International Workshop on Data Mining for Online Advertising}}
  \emph{(\bibinfo{series}{ADKDD '13})}. \bibinfo{publisher}{ACM},
  \bibinfo{address}{New York, NY, USA}, Article \bibinfo{articleno}{1},
  \bibinfo{numpages}{9}~pages.
\newblock
\showISBNx{978-1-4503-2323-9}
\urldef\tempurl%
\url{https://doi.org/10.1145/2501040.2501979}
\showDOI{\tempurl}


\bibitem[\protect\citeauthoryear{Murali, Li, Mazzoleni, and Vaculin}{Murali
  et~al\mbox{.}}{2015}]%
        {budget_allocation}
\bibfield{author}{\bibinfo{person}{Pavankumar Murali}, \bibinfo{person}{Ying
  Li}, \bibinfo{person}{Pietro Mazzoleni}, {and} \bibinfo{person}{Roman
  Vaculin}.} \bibinfo{year}{2015}\natexlab{}.
\newblock \showarticletitle{Optimal Budget Allocation Strategies for Real Time
  Bidding in Display Advertising}. In \bibinfo{booktitle}{\emph{Proceedings of
  the 2015 Winter Simulation Conference}} \emph{(\bibinfo{series}{WSC '15})}.
  \bibinfo{publisher}{IEEE Press}, \bibinfo{address}{Piscataway, NJ, USA},
  \bibinfo{pages}{3146--3147}.
\newblock
\showISBNx{978-1-4673-9741-4}
\urldef\tempurl%
\url{http://dl.acm.org/citation.cfm?id=2888619.2889011}
\showURL{%
\tempurl}


\bibitem[\protect\citeauthoryear{Ostrovsky and Schwarz}{Ostrovsky and
  Schwarz}{2011}]%
        {ostrovsky}
\bibfield{author}{\bibinfo{person}{Michael Ostrovsky} {and}
  \bibinfo{person}{Michael Schwarz}.} \bibinfo{year}{2011}\natexlab{}.
\newblock \showarticletitle{Reserve Prices in Internet Advertising Auctions: A
  Field Experiment}. In \bibinfo{booktitle}{\emph{Proceedings of the 12th ACM
  Conference on Electronic Commerce}} \emph{(\bibinfo{series}{EC '11})}.
  \bibinfo{publisher}{ACM}, \bibinfo{address}{New York, NY, USA},
  \bibinfo{pages}{59--60}.
\newblock
\showISBNx{978-1-4503-0261-6}
\urldef\tempurl%
\url{https://doi.org/10.1145/1993574.1993585}
\showDOI{\tempurl}


\bibitem[\protect\citeauthoryear{{Roels} and {Fridgeirsdottir}}{{Roels} and
  {Fridgeirsdottir}}{2009}]%
        {roels}
\bibfield{author}{\bibinfo{person}{G. {Roels}} {and} \bibinfo{person}{K.
  {Fridgeirsdottir}}.} \bibinfo{year}{2009}\natexlab{}.
\newblock \showarticletitle{{Dynamic Revenue Management for Online Display
  Advertising}}.
\newblock  (\bibinfo{year}{2009}).
\newblock


\bibitem[\protect\citeauthoryear{Wang and Chen}{Wang and Chen}{2012}]%
        {wang}
\bibfield{author}{\bibinfo{person}{Jun Wang} {and} \bibinfo{person}{Bowei
  Chen}.} \bibinfo{year}{2012}\natexlab{}.
\newblock \showarticletitle{Selling Futures Online Advertising Slots via Option
  Contracts}. In \bibinfo{booktitle}{\emph{Proceedings of the 21st
  International Conference on World Wide Web}} \emph{(\bibinfo{series}{WWW '12
  Companion})}. \bibinfo{publisher}{ACM}, \bibinfo{address}{New York, NY, USA},
  \bibinfo{pages}{627--628}.
\newblock
\showISBNx{978-1-4503-1230-1}
\urldef\tempurl%
\url{https://doi.org/10.1145/2187980.2188160}
\showDOI{\tempurl}


\bibitem[\protect\citeauthoryear{Yuan, Wang, Chen, Mason, and Seljan}{Yuan
  et~al\mbox{.}}{2014}]%
        {yuan}
\bibfield{author}{\bibinfo{person}{Shuai Yuan}, \bibinfo{person}{Jun Wang},
  \bibinfo{person}{Bowei Chen}, \bibinfo{person}{Peter Mason}, {and}
  \bibinfo{person}{Sam Seljan}.} \bibinfo{year}{2014}\natexlab{}.
\newblock \showarticletitle{An Empirical Study of Reserve Price Optimisation in
  Real-time Bidding}. In \bibinfo{booktitle}{\emph{Proceedings of the 20th ACM
  SIGKDD International Conference on Knowledge Discovery and Data Mining}}
  \emph{(\bibinfo{series}{KDD '14})}. \bibinfo{publisher}{ACM},
  \bibinfo{address}{New York, NY, USA}, \bibinfo{pages}{1897--1906}.
\newblock
\showISBNx{978-1-4503-2956-9}
\urldef\tempurl%
\url{https://doi.org/10.1145/2623330.2623357}
\showDOI{\tempurl}


\end{thebibliography}

\end{document}